\documentclass[aps,prd,preprint,superscriptaddress,nofootinbib]{revtex4-1}
\usepackage{chay,graphicx}

\newcommand{\eps}{\epsilon}

\newcommand{\euv}{\epsilon_{\mathrm{UV}}}
\newcommand{\eir}{\epsilon_{\mathrm{IR}}}

\begin{document}

\title{Factorization of the dijet cross section with the Georgi jet algorithm in $e^+ e^-$ annihilation}

\author{Junegone Chay}
\email[E-mail:]{chay@korea.ac.kr}
 \affiliation{Department of Physics, Korea University, Seoul 136-713, Korea}
\author{Inchol Kim}
\email[E-mail:]{vorfeed@korea.ac.kr}
\affiliation{Department of Physics, Korea University, Seoul 136-713, Korea}

\begin{abstract} \vspace{0.1cm}\baselineskip 3.0 ex 
We consider the dijet cross section in $e^+ e^-$ annihilation using the Georgi jet algorithm, or the maximizing jet algorithm. The cross section
is factorized into the hard, collinear and soft parts. Each factorized function is computed to next-to-leading order, and is shown to be infrared finite.
The large logarithms are resummed at next-to-leading logarithmic accuracy.  By analyzing the phase space for the jet algorithm, the Georgi  
algorithm turns out to be equivalent to the Sterman-Weinberg and the  cone-type algorithms.
 
\end{abstract}

\maketitle

\baselineskip 3.0 ex

\section{Introduction}
The study of jets is essential in understanding the interwoven effects of strong interaction and in extracting the information on Standard Model
or beyond. The strong interaction is responsible for the collective behavior in forming jets, starting from the scattering of the colored partons to
the formation of hadrons, and subsequently into the collimated beams of hadrons, which are called jets. In order to describe jets, there should be an
appropriate jet algorithm which combines adjacent final-state particles such that infrared (IR) safety is guaranteed.

There are many jet algorithms in different types of scattering like $e^+ e^-$ annihilation or $pp$ scattering   \cite{Salam:2009jx}.
Recently a jet algorithm has been suggested by maximizing a given function for a jet \cite{Georgi:2014zwa}. It is basically proposed for $e^+ e^-$ 
annihilation, and this jet algorithm has been extended to hadron-hadron collisions \cite{Ge:2014ova,Bai:2014qca}.  
In this letter, we present the complete analysis  employing the soft-collinear effective theory (SCET) \cite{Bauer:2000ew,Bauer:2000yr,Bauer:2001yt}.

Once a jet algorithm is selected, it is crucial to see if the jet cross section can be factorized, and each factorized part  is IR finite.
  It has been known that not all the jet algorithms satisfy the factorization theorem \cite{Chay:2015dva}.  We systematically analyze 
the Georgi jet algorithm to show that it factorizes the dijet cross section in $e^+ e^-$ annihilation, and each factorized part is infrared finite.
In proving the IR safety, we use the dimensional regularization with the spacetime dimension $D=4-2\eps$ regulating both ultraviolet (UV) and IR
divergences. In this case, the dimensional regularization states that 
\begin{equation}
\mu^{\eps}\int_0^{\infty} dl\, l^{-1-\eps} = \frac{1}{\euv} -\frac{1}{\eir},
\end{equation}
where $l$ is a momentum variable. If we do not distinguish the UV and IR poles, the above integral vanishes since the integral is a scaleless 
integral. However, we distinguish the UV and IR poles here to identify the
sources of the divergence explicitly. We also employ the 
$\overline{\mathrm{MS}}$ scheme with $4\pi \mu_{\overline{\mathrm{MS}}}^2
= \mu^2 e^{\gamma_{\mathrm{E}}}$.

\section{Jet algorithm}
An iterative jet algorithm suggested by Georgi is to assign a function $G(P)$, where $P$ is the four-momentum of the collection of the particles
to be included in a jet. It is given by 
\begin{equation} \label{defjet}
G_{\beta} (P) = P^0 -\beta \frac{P^2}{P^0},  \ \ \mathrm{for} \ \ \beta>1,
\end{equation}
and we find the set $\alpha$ with the maximum value of $G$, with
\begin{equation}
P^{\mu} = \sum_{j\in \alpha} p_j^{\mu}.
\end{equation}
Intuitively, the function $G(p)$ becomes maximum when the particles are selected such that the jet has a larger energy, and 
simultaneously a smaller invariant mass. 
 
In SCET, the four momentum of a collinear particle in the lightlike $n$ direction scales as 
$p^{\mu} = (\overline{n}\cdot p, p_{\perp}, n\cdot p) \sim Q(1,\lambda,\lambda^2)$,
where $Q$ is the center-of-mass energy and $\lambda$ is a small parameter in SCET, with $n^2 = \overline{n}^2 =0$, and $n\cdot \overline{n} =2$.
 In order for the two terms in $G(P)$ to compete with each other,
they have to be of the same order. Since $P^0 \sim \mathcal{O}(1)$ and $P^2\sim \mathcal{O}(\lambda^2)$, $\beta$ is of order $1/\lambda^2 \gg 1$. 

\begin{figure}[b] 
\begin{center}
\includegraphics[width=5cm]{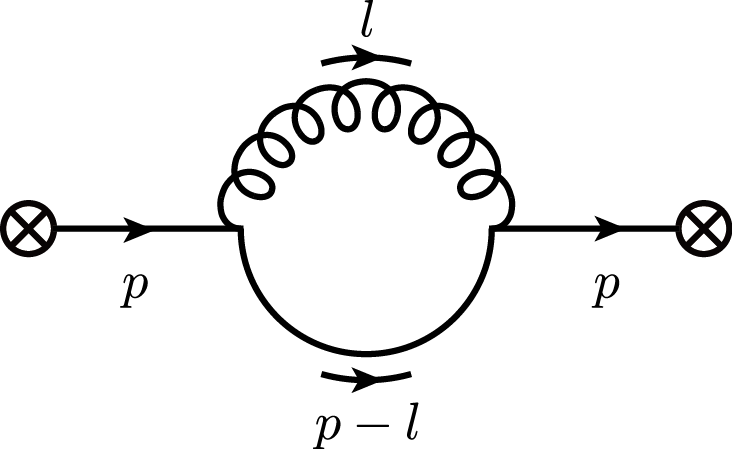}
\end{center}  
\vspace{-0.5cm}
\caption{\baselineskip 3.0ex  Particle configuration and the momentum assignment in constructing the phase space.\label{conf}}
\end{figure}

We consider the next-to-leading order (NLO) in which there are at most two particles in a jet. In this case, the backbone process is the 
production of a quark-antiquark pair with a virtual or real gluon. This contribution is 
obtained by cutting the diagram in  Fig.~\ref{conf}, which corresponds to the matrix elements squared for the jet cross
section.  If a single line is cut,  it yields the virtual correction. When we cut the loop, there are two final-state particles  
with momenta $l$ (for a gluon) and $p-l$ (for a quark).  A nontrivial jet algorithm results from the jet with two particles in it,
and we consider the kinematic constraint from the jet algorithm.

The collinear jet momentum $p^{\mu}$ in the $n$ direction $p^{\mu} =  (p_-,  p_{\perp}, p_+) \sim 
Q (1,\lambda, \lambda^2)$, and we choose the jet direction such that $\mathbf{p}_{\perp}=0$, and $p_- =Q$. 
The collinear gluon momentum $l^{\mu}$ scales as
$l^{\mu} = (l_- ,  l_{\perp},  l_+) \sim Q (1,\lambda, \lambda^2)$.
With this scaling behavior, the collinear momenta of the quark and the gluon can be written as
\begin{equation}
p_q^{\mu} = (Q-l_-, - l_{\perp}, p^2/Q -l_+), \ p_g^{\mu} = (l_-,  l_{\perp}, l_+),
\end{equation}
with their energies
\begin{equation}
E_q = \frac{1}{2} (Q-l_- +p^2/Q -l_+), \ E_g = \frac{1}{2} (l_- + l_+). 
\end{equation}
And the invariant-mass squared $p^2$ is given by
\begin{equation}
p^2 = (p_q + p_g)^2 = \frac{Ql_+}{1-l_-/Q}.
\end{equation}

For two particles inside a jet,  the criterial function in Eq.~(\ref{defjet}) for the two particles should be written as
\begin{equation} \label{jetalg}
G_{\beta} (p) > \mathrm{max}. \Bigl[ G_{\beta} (l), G_{\beta} (p-l)\Bigl],
\end{equation}
which can be expressed as
 \begin{eqnarray} \label{coljet}
&&l_+ < \frac{l_-}{4\beta} \Bigl(1-\frac{l_-}{Q}\Bigr), \ \ (0<l_- < \frac{Q}{2}), \nonumber \\
&&l_+ < \frac{Q}{4\beta}\Bigl(1-\frac{l_-}{Q}\Bigr)^2, \ \  (\frac{Q}{2} <l_- <Q).
\end{eqnarray}
 

In order to avoid double counting, we subtract the contribution corresponding to soft mode $l^{\mu} = Q(\lambda^2, \lambda^2, \lambda^2)$. 
This process is referred to as the zero-bin subtraction \cite{Manohar:2006nz}. The phase space for the zero-bin contribution to leading order 
in $\lambda$ is given by
\begin{equation}
l_+ < \frac{l_-}{4\beta}.  
\end{equation}

For the soft part, the phase space constraint from the jet algorithm in Eq.~(\ref{jetalg}) for the soft gluon is written as
\begin{eqnarray}
&& l_+ < \frac{l_-}{4\beta} , \ \ (n \ \mathrm{jet}),  \nonumber\\
&&  l_- < \frac{l_+}{4\beta} , \ \ (\overline{n} \ \mathrm{jet}),  \nonumber\\
&& l_- +l_+ < 2\delta Q, \ (\mathrm{jet} \ \mathrm{veto}).
\end{eqnarray}
In the first two constraints for $n$ and $\overline{n}$ jets, the denominator is actually $4\beta-1$. But to leading order in $\lambda$, it is 
replaced by $4\beta = 1/b$ since $\beta \sim \lambda^{-2}$. Here $b = 1/4\beta$ is a small parameter of order $\lambda^2$.

Note that there is an additional constraint for the jet veto. We introduce the quantity $\delta$
such that the energy fraction of the soft particle outside the jet should be less than $\delta$. It is not explicitly stated in the original jet algorithm, but
this jet veto is needed to render the soft function IR finite. From the power counting, we also require that
$\delta \sim \mathcal{O} (\lambda^2)$.

\section{Factorization of the dijet cross section}
The dijet cross section is given as \cite{Chay:2015ila}
\begin{equation} \label{facjet} 
\sigma_{\mathrm{jet}} =\sigma_0  H(Q^2,\mu)    \mathcal{J}_{n,\Theta} ( \mu)   \mathcal{J}_{\overline{n},\Theta} (\mu) 
\mathcal{S}_{\Theta} (\mu).
\end{equation}
Here $Q^2$ is the invariant-mass squared of the $e^+ e^-$ system, and $\sigma_0$ is the Born cross section for a given flavor $f$ 
of the quark-antiquark pair with the electric charge $Q_f$, given by
\begin{equation}
\sigma_0 = \frac{4\pi \alpha^2 Q_f^2 N_c}{3Q^2}.
\end{equation}
$H(Q^2,\mu)$ is the hard function which is obtained from the matching of the electromagnetic current between the full QCD and 
SCET at leading order as
\begin{equation}
\label{current}
J^{\mu} = C(Q^2,\mu) \overline{\chi}_n \tilde{Y}_n^{\dagger} \gamma^{\mu} \tilde{Y}_{\overline{n}} \chi_{\overline{n}},
\end{equation}
and $H(Q^2, \mu) = |C(Q^2,\mu)|^2$. To one loop, it is given by \cite{Manohar:2003vb}
\begin{equation} \label{hard}
H(Q^2,\mu) = 1+\frac{\alpha_s C_F}{2\pi}
\Bigl (-\ln^2  \frac{\mu^2}{Q^2} -3 \ln \frac{\mu^2}{Q^2} -8
+\frac{7\pi^2}{6}\Bigr). 
\end{equation}
And $\chi_n$ is a gauge-invariant collinear quark with a collinear Wilson line $\chi_n = W_n^{\dagger} \xi_n$, and $\tilde{Y} (x)$ is the soft
Wilson line \cite{Chay:2004zn} 
\begin{equation}
\label{tsoft} 
\tilde{Y}_n (x) = \mathrm{P} \exp\Bigl[ig\int^{\infty}_x ds n\cdot A_s (sn)\Bigr]_, 
\end{equation}  
where `P' denotes the path ordering.

The unintegrated jet function is defined as 
\begin{equation}
\sum_{X_n} \langle 0| \chi_n^{\alpha} |X_n\rangle \Theta_J \langle X_n | \overline{\chi}_n^{\beta}|0\rangle =
\int \frac{d^4 p_{X_n}}{(2\pi)^3} \frac{\FMslash{n}}{2} \overline{n} \cdot p_{X_n} J_{n,\Theta}  (p_{X_n}^2,\mu) \delta^{\alpha\beta},
\end{equation}
where $\Theta_J $ is the constraint specified by the jet algorithm, and $|X_n\rangle$ is the state for the collinear particles in the $n$ direction. 
The integrated jet function $\mathcal{J}_{n,\Theta}$ is obtained 
from the unintegrated jet function as
\begin{equation} \label{inunin}
\mathcal{J}_{n,\Theta}  (\mu)= \int dp^2 J_{n,\Theta} (p^2,\mu).
\end{equation}
The soft function $\mathcal{S}_{\Theta}$ with the jet algorithm is given as
  \begin{equation}
\mathcal{S}_{\Theta} = \sum_{X_s} \frac{1}{N_c} \mathrm{Tr} \langle 0| \tilde{Y}_{\overline{n}}^{\dagger}
\tilde{Y}_n |X_s\rangle \Theta_{\mathrm{soft}} \langle X_s |\tilde{Y}_n^{\dagger} \tilde{Y}_{\overline{n}} |0\rangle,
\end{equation}
where $\Theta_{\mathrm{soft}}$ dictates the jet algorithm for the soft particles, and $|X_s\rangle$ is the state for soft particles. The jet and the soft
functions are computed to next-to-leading order.

\begin{figure}[b] 
\begin{center}
 \includegraphics[width=16cm]{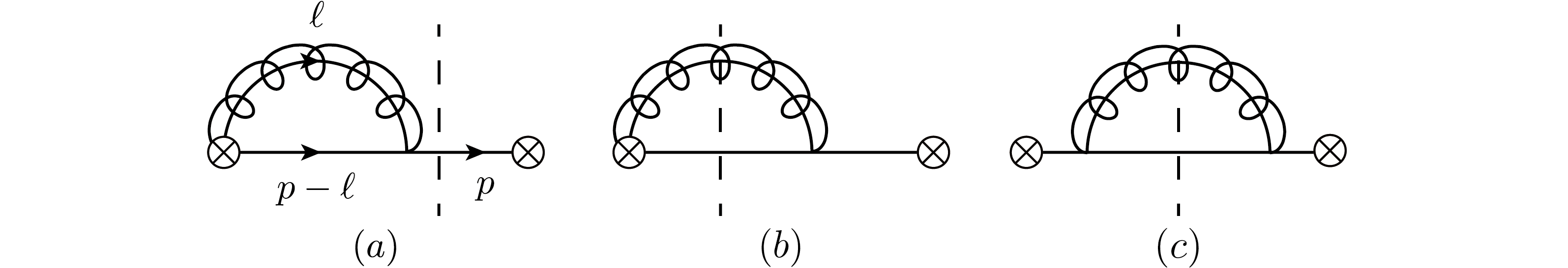}
\end{center}  
\vspace{-0.3cm}
\caption{\baselineskip 3.0ex  Feynman diagrams for the jet  function at one loop  
(a) virtual correction (b) real gluon emission from the Wilson line (c) real gluon emission.\label{intjet}}
\end{figure}
 
\section{Jet function}
The Feynman diagrams for the jet function at order $\alpha_s$ is shown in Fig.~\ref{intjet}, with the mirror images omitted for (a) and (b).
The dashed line represents the cut. Figure~\ref{intjet} (a) is the virtual correction, and is unaffected by the jet algorithm. The naive collinear
contribution $\tilde{M}_a$, the zero-bin contribution $M_a^0$, and the net collinear contribution $M_a = \tilde{M}_a -M_a^0$ are given by
\begin{eqnarray}
\tilde{M}_a &=& \frac{\alpha_s C_F}{2\pi} \Bigl( \frac{1}{\euv} -\frac{1}{\eir}\Bigr) \Bigl( \frac{1}{\eir} +1 
+\ln \frac{\mu}{Q}\Bigr), 
\nonumber \\
M_a^0 &=& -\frac{\alpha_s C_F}{2\pi} \Bigl( \frac{1}{\euv} -\frac{1}{\eir}\Bigr)^2,  \nonumber \\
M_a &=&  \tilde{M}_a - M_a^0 = \frac{\alpha_s C_F}{2\pi} = \frac{\alpha_s C_F}{2\pi} 
\Bigl( \frac{1}{\euv} -\frac{1}{\eir}\Bigr) \Bigl( \frac{1}{\euv} +1 +\ln \frac{\mu}{Q}\Bigr).
\end{eqnarray}

Figure~\ref{intjet} (b) and (c) represent the 
real gluon emissions.  The naive collinear contribution from Fig.~\ref{intjet} (b) is given as
\begin{equation}
\tilde{M}_b =
\frac{\alpha_s C_F}{2\pi} \Bigl[ \frac{1}{2\eir^2} +\frac{1}{\eir}\Bigl( 1+ \frac{1}{2}
\ln \frac{\mu^2}{bQ^2} \Bigr)  
+ \ln \frac{\mu^2}{bQ^2} +\frac{1}{4} \ln^2 \frac{\mu^2}{bQ^2}
+3 +\ln 2 -\frac{7}{24}\pi^2\Bigr], 
\end{equation} 
while the zero-bin contribution is given as
\begin{eqnarray}
M_b^0 &=& 
\frac{\alpha_s C_F}{2\pi} \frac{1}{2}\Bigl( \frac{1}{\euv}-\frac{1}{\eir}\Bigr) \Bigl( 
\frac{1}{\euv} -\frac{1}{\eir} +\ln b\Bigr).
\end{eqnarray}
The net collinear contribution is given by
\begin{eqnarray}
M_b &=& \tilde{M}_b -M_b^0 \nonumber \\
&=& \frac{\alpha_s C_F}{2\pi} \Bigl[ -\frac{1}{2\euv^2} +\frac{1}{\euv\eir} + \frac{1}{\eir}
\Bigl( 1+\ln \frac{\mu}{Q}\Bigr)  -\frac{1}{2\euv} \ln b   \nonumber \\
&&+ \ln \frac{\mu^2}{bQ^2} +\frac{1}{4} \ln^2 \frac{\mu^2}{bQ^2}
+3 +\ln 2 -\frac{7}{24}\pi^2\Bigr]. 
\end{eqnarray}

The naive collinear contribution from Fig.~\ref{intjet} (c) is given as
\begin{eqnarray}
M_c &=&
\frac{\alpha_s C_F}{2\pi}  \Bigl( -\frac{1}{2\eir} -\frac{1}{2} \ln \frac{\mu^2}{bQ^2} -1 -\frac{1}{2} \ln 2\Bigr).
\end{eqnarray}
The zero-bin contribution is suppressed and neglected. Including the wave function renormalization and the residue at one loop
\begin{equation}
Z_{\xi}^{(1)} +R_{\xi}^{(1)} = \frac{\alpha_s C_F}{2\pi} \Bigl( \frac{1}{2\eir} -\frac{1}{2\euv}\Bigr),
\end{equation}
the collinear contribution at order $\alpha_s$ is given by
\begin{eqnarray}
M_{\mathrm{coll}} &=& 2(M_a +M_b) + M_c +Z_{\xi}^{(1)} +R_{\xi}^{(1)}  \nonumber \\
&=& \frac{\alpha_s C_F}{2\pi}  \Bigl[ \frac{1}{\euv^2 }+\frac{1}{\euv} \Bigl( \frac{3}{2} + \ln \frac{\mu^2}{bQ^2} \Bigr)  \nonumber \\
&&+\frac{3}{2} 
\ln \frac{\mu^2}{bQ^2} +\frac{1}{2} \ln^2 \frac{\mu^2}{bQ^2}  +5+\frac{3}{2}\ln 2  -\frac{7}{12}\pi^2 \Bigr].    
\end{eqnarray}
Note that the collinear part is IR finite, and it only contains the UV divergence. After removing the UV divergence, the collinear jet function 
at one loop is given by
\begin{equation}
\mathcal{J}_n^{(1)} (Q,b,\mu) = \frac{\alpha_s C_F}{2\pi}  \Bigl(\frac{3}{2} 
\ln \frac{\mu^2}{bQ^2} +\frac{1}{2} \ln^2 \frac{\mu^2}{bQ^2}  +5+\frac{3}{2}\ln 2  -\frac{7}{12}\pi^2 \Bigr),
\end{equation}
from which the anomalous dimension of the jet function is obtained as
\begin{equation}
\gamma_J =  \frac{\alpha_s C_F}{2\pi} \Bigl( 2\ln \frac{\mu^2}{bQ^2} +3\Bigr).
\end{equation}

\section{Soft function}

\begin{figure}[b] 
\begin{center}
\includegraphics[width=16cm]{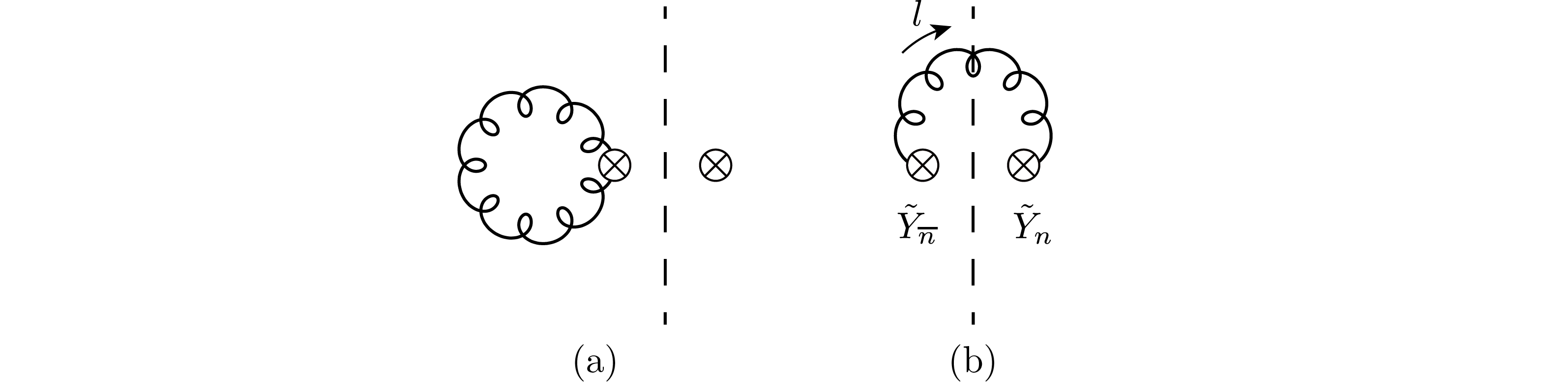}
\end{center}  
\vspace{-0.3cm}
\caption{\baselineskip 3.0ex Feynman diagrams for the soft  function at one loop  
(a) virtual corrections (b) real gluon emission.\label{feynsoft}}
\end{figure}

The Feynman diagrams for the soft function at one loop are shown in Fig.~\ref{feynsoft}, where the hermitian conjugates are omitted. The soft part
is given as
\begin{equation}
M_{\mathrm{soft}}= \frac{\alpha_s C_F}{2\pi} \Bigl(\frac{2}{\euv} \ln b + 4\ln b \ln \frac{\mu}{2\delta Q} -\ln^2 b -\frac{\pi^2}{3}\Bigr),
\end{equation}
and it is also IR finite. The soft function at one loop is given by
\begin{equation}
\mathcal{S}_{\Theta}^{(1)} = \frac{\alpha_s C_F}{2\pi} \Bigl(  4\ln b \ln \frac{\mu}{2\delta Q} -\ln^2 b -\frac{\pi^2}{3}\Bigr).
\end{equation}
Also the anomalous dimension for the soft function is obtained as
\begin{equation}
\gamma_S = \frac{2\alpha_s C_F}{\pi} \ln b.
\end{equation}

\section{Resummed dijet cross section}

Combining the hard function to one loop given by
\begin{equation} \label{hard}
H(Q^2,\mu) = 1+\frac{\alpha_s C_F}{2\pi}
\Bigl (-\ln^2  \frac{\mu^2}{Q^2} -3 \ln \frac{\mu^2}{Q^2} -8
+\frac{7\pi^2}{6}\Bigr),
\end{equation}
with the jet and soft functions,  the dijet cross section at NLO is given by
\begin{equation} \label{nloxsec}
\sigma^{(1)} = \sigma_0 \frac{\alpha_s C_F}{2\pi} \Bigl( -(4  \ln 2\delta +3)\ln b +2+3\ln 2 -\frac{\pi^2}{3}\Bigr).
\end{equation}
 
Since the cross section involves the factorized parts which contain large logarithms, the cross section should be resummed for 
the large logarithm, and it is obtained by solving
the renormalization group equation at next-to-leading logarithmic (NLL) accuracy here.  The anomalous dimensions of the hard, jet and soft functions can 
be cast into the form
\begin{eqnarray}
\gamma_H &=& \Gamma_{\mathrm{cusp}} (\alpha_s) \ln \frac{Q^2}{\mu^2} +\Gamma^H (\alpha_s), \nonumber \\
\gamma_J &=& -\frac{1}{2} \Gamma_{\mathrm{cusp}} (\alpha_s) \ln \frac{\mu_J^2}{\mu^2} +\Gamma^J (\alpha_s), \nonumber \\
\gamma_S &=& \Gamma^S (\alpha_s),
\end{eqnarray}
where $\mu_J = \sqrt{b} Q$ is the jet scale, and $\Gamma_{\mathrm{cusp}}$ is the cusp anomalous dimension of the hard function. 
It can be explicitly verified that $\gamma_H + 2\gamma_J + \gamma_S=0$, which implies that the jet cross section is independent of the 
renormalization scale $\mu$. To NLL order,
we need the cusp anomalous dimension to two loop order, the remaining anomalous dimensions $\Gamma^H$, $\Gamma^J$, $\Gamma^S$ to 
one loop order, and the hard, jet and soft functions at tree level.

The renormalization group equation for the hard, jet and soft functions is of the form
\begin{equation}
\frac{d}{d\ln \mu} f_i (\omega_i,\mu) = \Bigl[a_i (\alpha_s) \ln \frac{\omega_i^2}{\mu^2} + b_i (\alpha_s)\Bigr] f_i (\omega_i,\mu),
\end{equation}
where $\omega_i = Q, \mu_J, \mu_S$ for $i=H, J, S$ with the soft scale $\mu_S = 2\delta Q$. 
And the solution is given by \cite{Becher:2006mr}
\begin{equation}
f_i (\omega_i,\mu) = \exp \Bigl[2 S_i (\mu_i,\mu) -B_i (\mu_i,\mu)\Bigr] \Bigl(\frac{\omega_i^2}{\mu_i^2}\Bigr)^{-A_i (\mu_i,\mu)} 
f_i (\omega_i,\mu),
\end{equation}
where $\mu_i$ are the factorization scales from which the hard, jet and soft functions are evolved to the renormalization scale $\mu$. 
To NLL order, $S_i$, $A_i$ and $B_i$ are given as
\begin{eqnarray}
S_i (\mu_i,\mu) &=& \frac{a_i^0}{4\beta_0^2} \Bigl[ \frac{4\pi}{\alpha_s (\mu_i)} \Bigl( 1-\frac{\alpha_s (\mu_i)}{\alpha_s(\mu)}
-\ln \frac{\alpha_s (\mu)}{\alpha_s (\mu_i)} \Bigr)  +\frac{\beta_1}{2\beta_0} \ln^2 \frac{\alpha_s (\mu)}{\alpha_s (\mu_i)}  \nonumber \\
&& +\Bigl( \frac{a_i^1}{a_i^0}-\frac{\beta_1}{\beta_0} \Bigr) 
\Bigl( 1- \frac{\alpha_s (\mu_i)}{\alpha_s (\mu)}  +\ln \frac{\alpha_s (\mu)}{\alpha_s (\mu_i)} \Bigr)\Bigr], \nonumber \\
A_i (\mu_i,\mu) &=& \frac{a_i^0}{2\beta_0} \Bigl[ \ln \frac{\alpha_s (\mu)}{\alpha_s (\mu_i)} + \Bigl( \frac{a_i^1}{a_i^0} 
-\frac{\beta_1}{\beta_0} \Bigr) \frac{\alpha_s (\mu) -\alpha_s (\mu_i)}{4\pi} \Bigr], \nonumber \\
B_i (\mu_i,\mu) &=& \frac{b_i^0}{2\beta_0} \ln \frac{\alpha_s (\mu)}{\alpha_s (\mu_i)}.
\end{eqnarray}

The QCD $\beta$ function, the cusp anomalous dimension, and the anomalous dimensions  in the $\overline{\mathrm{MS}}$ scheme are given as
\begin{eqnarray}
&&\beta(\alpha_s) = \frac{d}{d\ln \mu} \alpha_s = -2 \alpha_s \Bigl[ \beta_0 \frac{\alpha_s}{4\pi} 
+ \beta_1 \Bigl(\frac{\alpha_s}{4\pi}\Bigr)^2 +\cdots \Bigr], \nonumber \\
&&\Gamma_{\mathrm{cusp}} (\alpha_s) = \Gamma_0 \frac{\alpha_s}{4\pi} +\Gamma_1 \Bigl(\frac{\alpha_s}{4\pi}\Bigr)^2+\cdots, \nonumber
\\
&&a_i (\alpha_s) =  a_i^0  \frac{\alpha_s}{4\pi} + a_i^1 \Bigl(\frac{\alpha_s}{4\pi}\Bigr)^2 +\cdots,  \ \ 
b_i (\alpha_s) = b_i^0  \frac{\alpha_s}{4\pi} +b_i^1 \Bigl(\frac{\alpha_s}{4\pi}\Bigr)^2 +\cdots,
\end{eqnarray}
where the expansion coefficients for the QCD $\beta$ function to two-loop order are 
\begin{equation}
\beta_0 = \frac{11}{3}C_A -\frac{4}{3} T_F n_f, \ \beta_1 = \frac{34}{3}C_A^2 -\frac{20}{3} C_A T_F n_f -4C_F T_F n_f.
\end{equation}
And the cusp anomalous dimension to two loop order are given as \cite{Korchemsky:1987wg,Korchemskaya:1992je}
\begin{equation}
\Gamma_0 = 8C_F,  \ \ \Gamma_1 = 8C_F\Bigl[\Bigl( \frac{67}{9} -\frac{\pi^2}{3}\Bigr) C_A -\frac{20}{9}T_F n_f\Bigr].
\end{equation}

\begin{figure}[t] 
\begin{center}
\includegraphics[width=16cm]{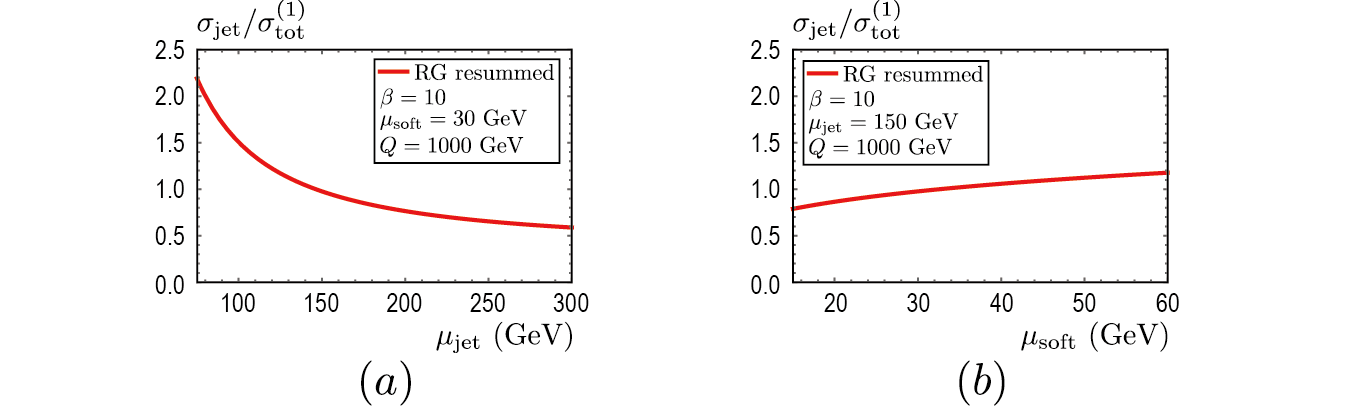}
\end{center}  
\vspace{-0.3cm}
\caption{\baselineskip 3.0ex The dependence of the jet cross section on the renormalization scales. (a) $\mu_J/2<\mu_{\mathrm{jet}} <2 \mu_J$ with
$\mu_{\mathrm{soft}}= 2\delta Q$ and $\mu_ H=Q$ fixed, (b) $\mu_S/2<\mu_{\mathrm{soft}} <2 \mu_S$ with
$\mu_{\mathrm{jet}}= \sqrt{b} Q$ and $\mu_ H=Q$ fixed.\label{xsecscale}}
\end{figure}

The dependence of the jet cross section on the jet scale $\mu_{\mathrm{jet}}$ and the soft scale $\mu_{\mathrm{soft}}$ is shown in 
Fig.~\ref{xsecscale}.  The hard scale is set at $\mu_H =Q$. In the first figure, the jet scale varies between $\mu_J/2$ and $2\mu_J$ 
where $\mu_J =\sqrt{b}Q$, while the soft scale is fixed. In the second figure, 
the soft scale varies between $\mu_S/2$ to $2\mu_S$ where $\mu_S = 2\delta Q$, while the jet scale is fixed. 
Since the parameter $\delta$ in the jet veto should be of order
$\lambda^2$, we put $2\delta \sim b$.  The dijet cross section is normalized to the total cross section at order $\alpha_s$, which is given by
\begin{equation}
\sigma_{\mathrm{tot}}^{(1)} = 1+\frac{\alpha_s}{\pi}.
\end{equation}
Then the ratio $\sigma_{\mathrm{jet}}/\sigma_{\mathrm{tot}}^{(1)}$ is the two-jet fraction.
In Fig.~\ref{xsecscale}, the jet cross section shows mild dependence on the jet and soft scales. 
Due to the different scales on the horizontal axes in the figure,
the dependence on the jet scale is actually milder than that on the soft scale. 

\begin{figure}[b] 
\begin{center}
\includegraphics[width=16cm]{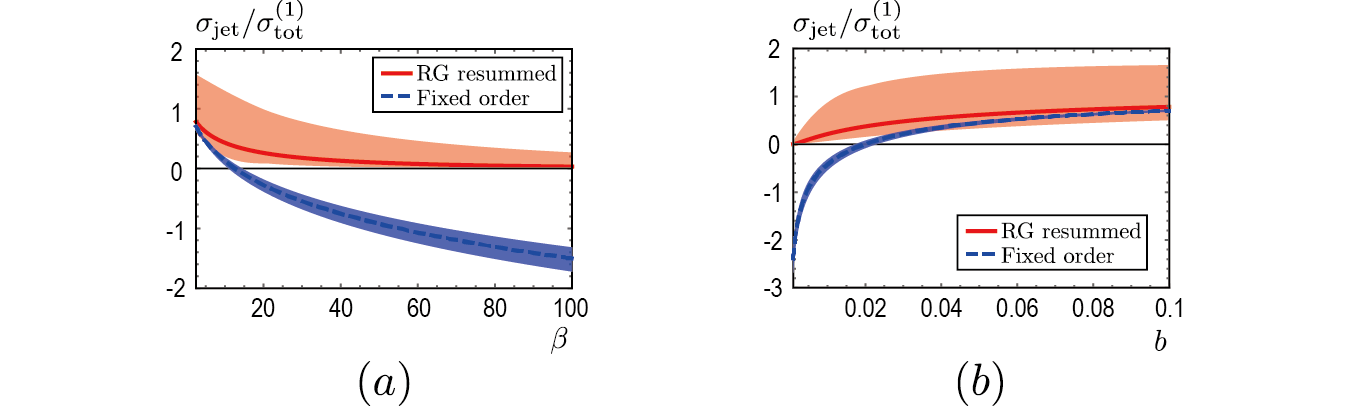}
\end{center}  
\vspace{-0.3cm}
\caption{\baselineskip 3.0ex The dijet fraction as a function of (a) $\beta$ (b) $b=1/4\beta$. The bands show the theoretical uncertainties. The solid
line in the resummed cross section is obtained with $\mu_{\mathrm{hard}}=Q$, 
$\mu_{\mathrm{jet}} = \mu_J$, $\mu_{\mathrm{soft}} = \mu_S$. The dashed line is the NLO cross section with $\mu=Q$. 
\label{xsecb}}
\end{figure}

In Fig.~\ref{xsecb}, the jet cross sections at NLL order and at NLO order are plotted with respect to the large parameter $\beta$, and the small 
parameter $b$ respectively. At NLL order, the theoretical uncertainty is obtained by varying the 
hard scale from $Q/2$ to $2Q$, the jet scale from $\mu_J/2$ to $2\mu_J$, and the soft scale from $\mu_S/2$ to $2\mu_S$. At NLO, all the scales 
are set to $\mu =\mu_H = \mu_J =\mu_S$, and the renormalization scale $\mu$ is varied  from $Q/2$ to $2Q$.  The solid line in the
band is for the scale $\mu_{\mathrm{hard}}=Q$, $\mu_{\mathrm{jet}} = \mu_J$ and $\mu_{\mathrm{soft}} = \mu_S$, and the dashed line is 
obtained by setting $\mu=Q$.

For large $\beta$ or small $b$, the fixed-order result becomes negative and the perturbative results lose physical meaning. On the other hand, 
the resummed result remains positive and is suppressed for $b\rightarrow 0$, while the fixed-order results diverges. Therefore  the dijet cross section
becomes meaningful only after the large logarithms are resummed.

\section{Conclusion}
We have shown that the dijet cross section in $e^+ e^-$ annihilation is factorized using SCET, in the sense that each factorized part is IR finite. 
In Ref.~\cite{Thaler:2015uja}, it is shown that the Georgi algorithm, jets based on the jettiness, and the cone algorithm are basically 
equivalent by introducing a meta function. We take a different approach, that is, the structure of the phase space for the collinear and soft parts 
constrained by the jet algorithm completely determines the characteristics of the jet algorithms including the infrared safety.
In this perspective, we can compare the characteristics of the Georgi jet algorithm with the existing jet algorithms. Compared to the 
results in Ref.~\cite{Chay:2015ila}, where the Sterman-Weinberg and the cone-type jet algorithms have been analyzed, the structure of the phase
space for the jet and the soft functions of these algorithms are basically the same as the Georgi algorithm. Therefore the structure of the 
divergence shows the similar behavior as well. As can be seen
in Eq.~(\ref{nloxsec}), the jet cross section is analogous to that in the Sterman-Weinberg algorithm \cite{Sterman:1977wj}, and the small parameter 
$b = 1/4\beta$ plays the role of the angular size in the Sterman-Weinberg or the cone-type jet algorithms.

In Ref.~\cite{Chay:2015dva},
the generalized exclusive $k_T$ algorithm is investigated. The divergence structure, or the shape of the phase spaces has been classified by 
specifying the parameter $\alpha$. The cone-type and the Sterman-Weinberg jet algorithms belong to the category with 
$\alpha<0$ along with the exclusive anti-$k_T$ algorithm.  From the shape of the phase space in the Georgi jet algorithm, we can conclude that
 it is kinematically similar to the exclusive $k_T$ algorithm with $\alpha<0$, in which an additional jet veto is needed in the soft part. 

We have resummed large logarithms appearing in the jet and soft functions. However, there is an issue on resumming the logarithms of the small
jet radius, which corresponds to logarithms of $b$ \cite{Becher:2015hka,Chien:2015cka}. It is claimed to be obtained by introducing additional
new degrees of freedom in SCET. But this is not pursued in this letter.

It will be interesting if all the features on the factorization property, the divergence structure and the shape of the phase space are sustained in
hadron-hadron scattering. All the cone-type and the inclusive $k_T$ jet algorithms fall into the category with $\alpha<0$, but it remains to be seen
if the results remain the same or not due to the kinematic difference between $e^+ e^-$ annihilation and the $pp$ scattering.

 \section*{Acknowledgment}

The authors are supported by Basic Science Research Program through the National Research Foundation of Korea (NRF) funded by 
the Ministry of Education(Grant No. NRF-2014R1A1A2058142).

\end{document}